\documentclass[%
    reprint,
    amssymb,
    amsmath,
    aip,
    rsi,
]{revtex4-2}

\usepackage{amssymb}
\usepackage{graphicx}
\usepackage{docs}
\usepackage{bm}
\usepackage[colorlinks=true,linkcolor=blue]{hyperref}

\usepackage{siunitx}
\usepackage{ulem}

\expandafter\ifx\csname package@font\endcsname\relax\else
\expandafter\expandafter
\expandafter\usepackage
\expandafter\expandafter
\expandafter{\csname package@font\endcsname}
\fi

\begin{document}

\title{Fast-field-cycling, ultralow-field nuclear magnetic relaxation dispersion}%

\author{Sven Bodenstedt}
\affiliation{ICFO -- Institut de Ci\`encies Fot\`oniques, The Barcelona Institute of Science and Technology, 08860 Castelldefels (Barcelona), Spain}
\author{Morgan W. Mitchell}
\affiliation{ICFO -- Institut de Ci\`encies Fot\`oniques, The Barcelona Institute of Science and Technology, 08860 Castelldefels (Barcelona), Spain}
\affiliation{ICREA -- Instituci\'{o} Catalana de Recerca i Estudis Avan\c{c}ats, 08010 Barcelona, Spain}
\author{Michael C. D. Tayler}
\email{michael.tayler@icfo.eu}
\affiliation{ICFO -- Institut de Ci\`encies Fot\`oniques, The Barcelona Institute of Science and Technology, 08860 Castelldefels (Barcelona), Spain}

\date{May 31\textsuperscript{st} 2021}%

\maketitle

\section*{KEYWORDS}
Molecular motion; Ultralow-field nuclear magnetic resonance (ULF NMR); Optically pumped magnetometers (OPMs); Relaxation; Porous materials. 

\section*{Abstract}
Optically pumped magnetometers (OPMs) based on alkali-atom vapors are ultra-sensitive devices for dc and low-frequency ac magnetic measurements.  Here, in combination with fast-field-cycling hardware and high-resolution spectroscopic detection, we demonstrate applicability of OPMs in quantifying nuclear magnetic relaxation phenomena.  Relaxation rate dispersion across the \si{\nano\tesla} to \si{\milli\tesla} field range enables quantitative investigation of extremely slow molecular motion correlations in the liquid state, with time constants $>$\SI{1}{\milli\second}, and insight into the corresponding relaxation mechanisms.  The 10-20~\si{\femto\tesla\per\sqrt\hertz} sensitivity of an OPM between 10 Hz and 5.5 \si{\kilo\hertz} \textsuperscript{1}H Larmor frequency suffices to detect magnetic resonance signals from $\sim$0.1 mL liquid volumes imbibed in simple mesoporous materials, or inside metal tubing, following nuclear spin prepolarization adjacent to the OPM.  High-resolution spectroscopic detection can resolve inter-nucleus spin-spin couplings, further widening the scope of application to chemical systems.  Expected limits of the technique regarding measurement of relaxation rates above \SI{100}{\per\second} are discussed.  

\section*{Introduction}

Nano-scale dynamic processes that occur on \si{\milli\second} to \si{\micro\second} time scales, such as protein folding, aqueous complexation and surface adsorption phenomena, are often probed using nuclear magnetic relaxation dispersion (NMRD) techniques\cite{Kimmich1979BMR1,Kimmichbook, Kimmich2004PNMRS}, in which field-dependent relaxation rates of nuclear spins are used to infer correlation times for molecular reorientation\cite{Schneider2007book,Deutch1968AdvOptMagnReson3} and diffusive transport.  Beyond fundamental interests, insights from NMRD such as surface fractal dimension and roughness provide models for industrial catalysis and petrology, where liquids are confined inside porous solids and molecular diffusion is restricted by surface geometry\cite{Bychuk1995PRL74} as well as adsorption\cite{Guo2020PetSci17}, and in medicine assist the design of molecular agents for relaxation-contrast magnetic resonance imaging (MRI)\cite{Waddington2020SciAdv6}.  Furthermore, if coupled with spectroscopic dispersion via chemical shifts or spin-spin couplings, the dynamics can be related to specific molecular functional groups, facilitating analyses of chemical mixtures and biological specimens\cite{Korb2017PNMRS}.  

\begin{figure}
	\includegraphics[width=\columnwidth]{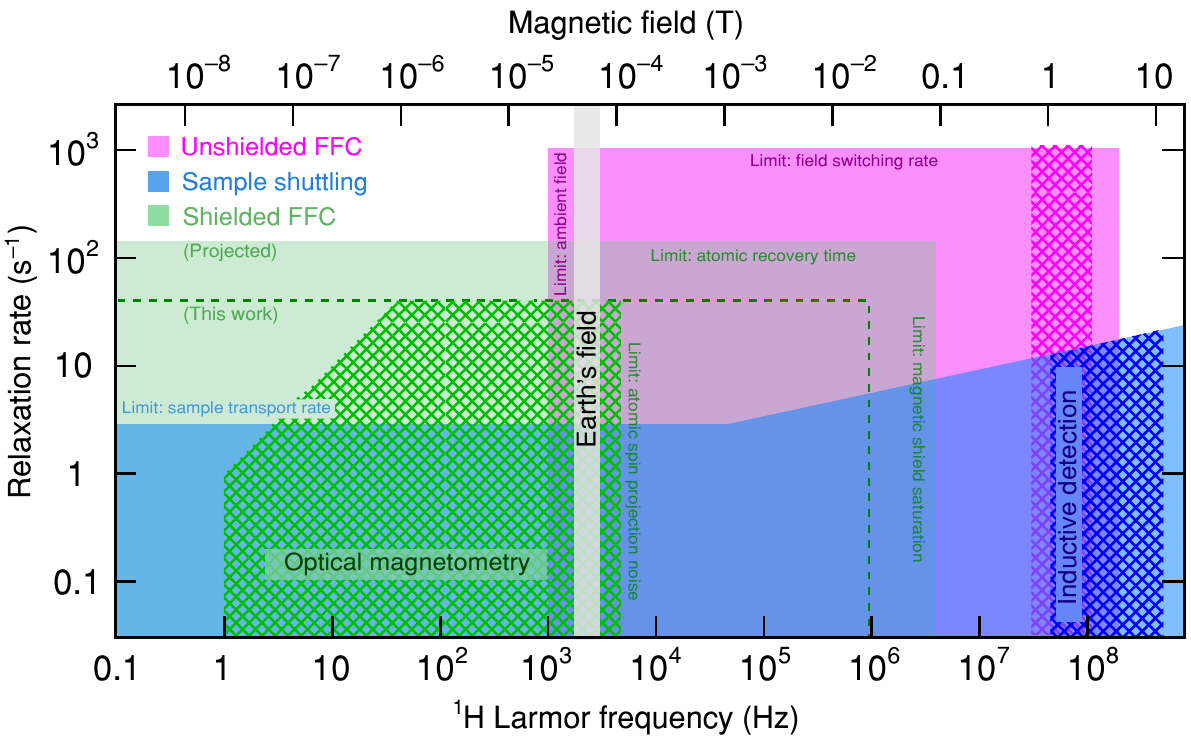}
	\caption{Limits of existing NMRD techniques.  Boundaries of the colored zones in the horizontal dimension indicate upper limits of longitudinal relaxation rate, ($T_{1,I}^{-1}$ \si{\per\second}), imposed by signal receiver dead time and/or speed of field-switching events.  Boundaries in the vertical dimension indicate limits to the range of magnetic fields achievable by electromagnetic sources.  Cross-hatching indicates Larmor frequencies where NMR signals are detected and corresponds to measurable transverse relaxation rates $T_{2,I}^{-1}$ (\si{\per\second}).  The diagonal of slope 1 corresponds to the spectroscopic resolution limit where rates of Larmor precession and transverse relaxation are equal: $B \gamma_I/(2\pi) = T_{2,I}^{-1}\sim T_{1,I}^{-1}$.}
	\label{fig:ExistingLimits}
\end{figure}

Accurate correlation times $\tau_{\rm c}$ can be obtained by measuring nuclear spin relaxation across a range of Larmor frequencies $B \gamma_I  \ll \tau_{\rm c}^{-1}$ to $B \gamma_I \gg  \tau_{\rm c}^{-1}$, where $B$ is the field strength and $\gamma_I$ is the nuclear gyromagnetic ratio.  Extremely slow correlations thus require measurements at ultralow magnetic fields within shielded enclosures such as a MuMetal~chamber.  The main existing NMRD technique uses fast-field-cycling (FFC) electromagnets\cite{Job1996RSI67,WardWilliams2018JPCC122,Ferrante2005AdvInChem57,Anoardo2001AMR20} of around \SI{1}{\tesla} for efficient inductive NMR signal detection, but these must be used unshielded with active cancelation of ambient fields to access below the geomagnetic field range\cite{Anoardo2003ApplMagnReson24,Kresse2011SSNMR40}.  Alternatively, NMRD is performed by transporting samples between persistent high- and ultralow-field locations\cite{Chou2012JMR214,Kaseman2020RSI91,Tayler2018JMR297,Ganssle2014ANIE53,Zhukov2018PCCP}, but relatively slow transport times limit the observable $\tau_{\rm c}$ at the high end.  The limits of these existing techniques are illustrated by the magenta- and blue-shaded regions, respectively, of \autoref{fig:ExistingLimits}.

In this work, we introduce a third scenario to address the the top-left portion of \autoref{fig:ExistingLimits} that lies outside the reach of inductive NMR pickup.  The speed of the FFC approach is combined with the low-frequency sensitivity of a  spin-exchange-relaxation-free (SERF)\cite{Happer1977PRA16,Allred2002PRL89,Savukov2005PRA71,Savukov2005PRL,Savukov2007JMR185,Ledbetter2008PRA77,Blanchard2016emagres,Tayler2017RSI91} optically pumped magnetometer (OPM) to perform NMRD at $^1$H Larmor frequencies from \SI{1}{\hertz} to \SI{10}{\kilo\hertz}, corresponding to the region of \autoref{fig:ExistingLimits} shaded in green.  The high sensitivity of SERF OPMs of order \SI{1}{\femto\tesla\per\sqrt\hertz}\cite{Budker2007NatPhys3,BudkerOpticalMagnetometry} at signal frequencies down to a few Hz, rivals the best Superconducting Quantum Interference Device (SQUID)\cite{McDermott2002Science295,Trahms2010MRI28} and high-Q inductive-pickup magnetometers below Earth's field\cite{Appelt2006NatPhys2,Suefke2015NatPhys11}, with the advantage of cryogen-free operation and simple tuning based on Hartmann-Hahn matching of the OPM and NMR spin ensembles.  In the ultralow-field NMRD context, the OPM is compatible both with MuMetal shielding and relatively weak prepolarizing fields of order \SI{10}{\milli\tesla}.  Magnetic fields for relaxation and detection are supplied accurately and precisely (within \SI{1}{\nano\tesla}) following a one-off calibration procedure and can be cycled in less than \SI{1}{\milli\second}, resulting in atomic-response-limited dead times. Based on this configuration, we are able to study spin relaxation phenomena that cannot be probed using conventional inductive field-cycling NMR procedures: (1) relaxation of liquids encased in metal tubing, demonstrated using aqueous solutions of paramagnetic impurities; (2) the full frequency dependence for motional correlations in a system of \textit{n}-octane (\textit{n}-C\textsubscript{8}$H$\textsubscript{18}) and \textit{n}-decane (\textit{n}-C\textsubscript{10}H\textsubscript{22}) absorbed in nanoscale confinement upon porous alumina and titania, as an example relevant to research in catalysis.  Results unambiguously support dynamics models involving molecular diffusion among paramagnetic sites on the pore surface; (3) chemical species resolution via spin-spin J couplings.

\subsection*{Diffusion dynamics from NMRD}
When it covers the appropriate range of fields and time scales, the relaxation measured in NMRD can relate model parameters of interest to molecular motion, surface structure and molecule-surface interactions \cite{Redfieldtheory,Kowalweskibook,Halle1998JMR135}. A central quantity of interest is the time correlation function $g(\tau) = \langle x(t) x(t+\tau)\rangle/\langle x(t) x(t)\rangle$ of the molecular motion.  This is related to observable relaxation by the quantity $j(\omega)=\int_0^\infty{g(\tau)\cos(\omega \tau)\mathrm{d}\tau}$, i.e. the cosine transform of $g(\tau)$.  Here we assume a simple but useful model where the local field is inhomogeneous, with a randomly oriented component of root-mean-square amplitude $B_\mathrm{rms}$.  The longitudinal relaxation rate is $[T_{1,I}(\omega_I)]^{-1} = \gamma_I^2 B_\mathrm{rms}^2 j(\omega_I)$ under standard perturbation (i.e.\ Redfield\cite{Redfieldtheory}) assumptions.  In the ideal case of unrestricted diffusion, a single correlation time is found, where $g(\tau)\propto \exp[-(\tau/\tau_{\rm c})]$ and the spectral density is a Lorentzian: $j(\omega)=\tau_{\rm c}/(1+\omega^2\tau_{\rm c}^2)$, where $\tau_{\rm c}$ is the characteristic diffusion time.  Inverse-square power law behavior is thus expected for $T_{1,I}^{-1}$ vs.\ $\omega_I$, for $\omega_I\gg\tau_{\rm c}^{-1}$.

Scenarios of constrained Brownian motion\cite{Klafter1986PNAS83} such as diffusion in pores may yield several concurrent dynamics modes.  Fitting to a distribution of correlation times $p(\tau')$ may be more appropriate: $[T_{1,I}(\omega_I)]^{-1} = \gamma_I^2 \int_{0}^{\infty} B_\mathrm{rms}^2 p(\tau_{\rm c}) \tau_{\rm c}/(1+\omega_I^2\tau_{\rm c}^2)\,\mathrm{d}\tau_{\rm c}$  \cite{Mcdonald2018PRE98,Miyaguchi2019PRE100}
, where $p(\tau_{\rm c})$ represents a probability distribution normalized to $\int_0^\infty p(\tau_{\rm c}) \, \mathrm{d} \tau_{\rm c} = 1$.  Kimmich and co-workers examine this approach to explain power-law relaxation behavior in porous glasses: $T_{1,I} \propto \omega_I^\xi$, where $0<\xi<2$\cite{Kimmich2002CP282,Zavada1999PRE59}.  Surface-induced relaxation is attributed to ``molecular reorientation mediated by translational displacement'' (RMTD), where diffusion across a rugged pore surface modulates intra-molecular spin-spin dipolar couplings and $p(\tau_{\rm c})$ is linked to the surface fractal dimension.  A breakdown of the power law at low frequencies $\omega_I$ indicates a maximum $\tau_{\rm c}$, which is connected to the longest distance a molecule can diffuse before leaving the surface phase or experiencing a different surface structure.  The value should depend on the molecule, due to different diffusion coefficients, as well as the porous medium. 

Moreover towards zero Larmor frequency, $T_{1,I}$ tends to a plateau defined by $[T_{1,I}(0)]^{-1} = \gamma_I^2  \int_{0}^{\infty} B_\mathrm{rms}^2 p(\tau_{\rm c}) \tau_{\rm c} = \gamma_I^2 B_\mathrm{rms}^2 \langle \tau_c \rangle$, where $\langle \tau_c \rangle$ is the mean correlation time\cite{Halle1998JMR135}.  This and the above measures all require $T_{1,I}$ to be known for frequencies below $\tau_{\rm c,max}^{-1}$, motivating ultra-low field measurement capability.  

\section*{Results}
\subsection*{Tunable NMR detection using optical magnetometry}
\label{sec:NMRDetection}
\begin{figure}[t]
	\includegraphics[width=\columnwidth]{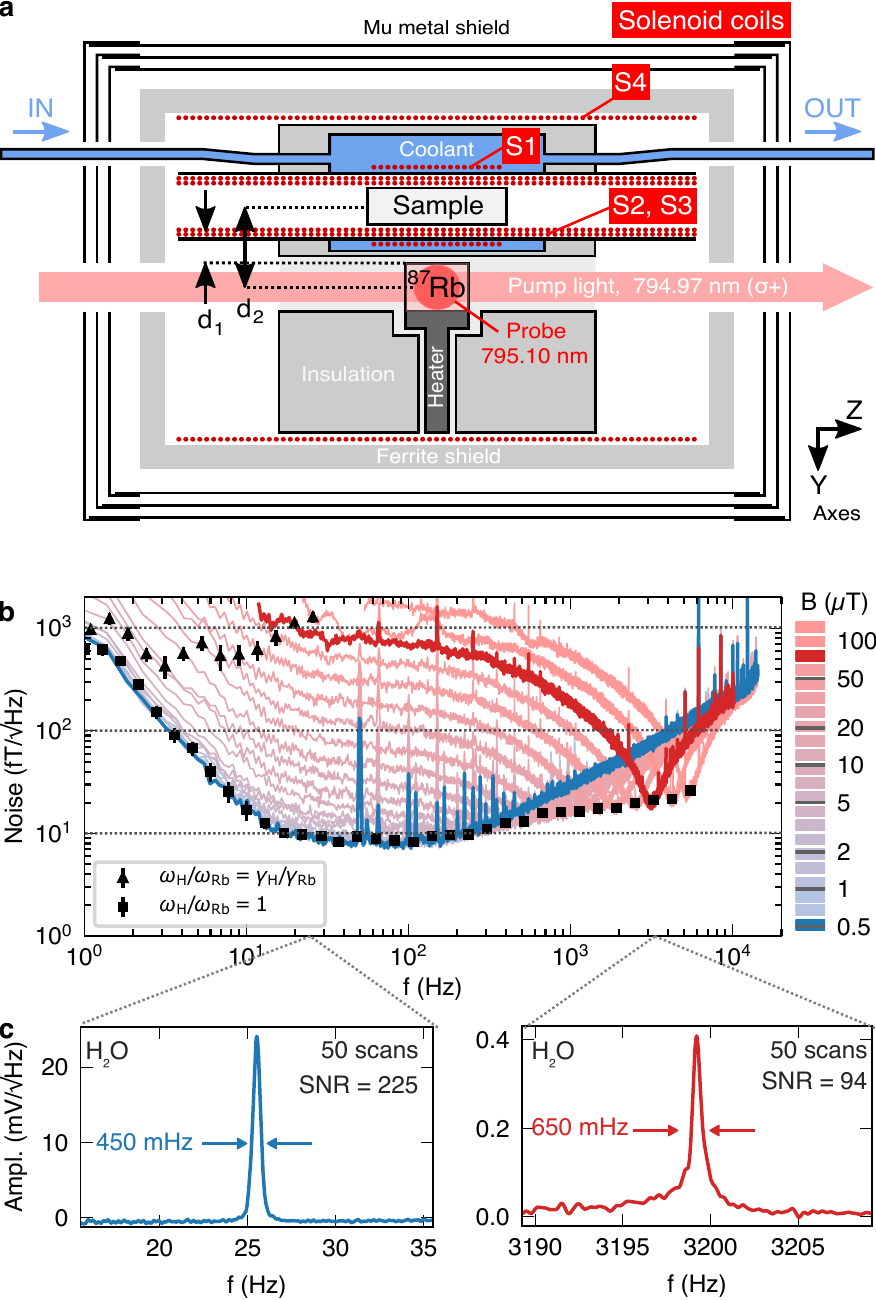}
	\caption{Tunable NMR detection via optically pumped magnetometer: (a) Schematic view of the apparatus through a vertical cross section.  Standoff distance between sample container atomic vapor cell $d_1 = 3.5\,\mathrm{mm}$.  Center-to-center distance between atomic cell and NMR sample $d_2 = 10\,\mathrm{mm}$; (b) Frequency dependence of magnetometer noise in field-equivalent units at bias fields $B_{z,I}$ from 0.1 to 100 $\mu$T.  Plot markers indicate sensitivity at the \textsuperscript{1}H Larmor frequency under the conditions of ($\blacksquare$) tuned and ($\blacktriangle$) untuned \textsuperscript{87}Rb Larmor frequencies; error bars correspond to the sensitivity's  root-mean-square deviation within a \SI{1}{Hz} window around the Larmor frequency. (c) Representative NMR signal and noise amplitudes in the tuned case for a sample of 1.8\,mL milli-Q water, measured after pre-polarization at 20\,mT and a transverse $\pi/2$ pulse. }\label{fig:OPMresponse}
\end{figure}

\autoref{fig:OPMresponse}a shows an experimental setup for FFC NMR with OPM detection.  A \SI{2}{\milli\liter} vial containing the NMR sample sits within four coaxial solenoids (S1 -- S4), which provide $z$-oriented fields.  Liquid coolant flows around the sample chamber, with S1 moreover immersed in the flowing coolant to maintain a sample temperature around \SI{30}{\celsius}. The short S1 solenoid is used to produce fields up to \SI{20}{\milli\tesla} to polarize the nuclear spins in the sample, while S2 and S3 provide weaker fields for Larmor precession.  Adjacent to this chamber is a heated glass cell containing \textsuperscript{87}Rb vapor, with optical access along $x$ and $z$ directions for probing and pumping of the alkali spin angular momentum $\bm{\mathrm{S}}$, respectively.    

A magnetic field $\bm{\mathrm{B}} = (B_1\cos{\omega t},B_1 \sin{\omega t}, B_{z,S})$ in Cartesian coordinates is assumed at the atoms, as the sum of a constant bias field $B_{z,S}$ along the $z$ axis and an effective rotating field $B_1$ induced by precession of nuclear magnetization in the NMR sample in the $xy$ plane.  Assuming the nuclei experience a field $B_{z,I}$ along the $z$ axis, then $\omega=\omega_I=\gamma_I B_{z,I}$, and $B_1$ is proportional to the amplitude of the nuclear magnetization. 

Dynamics of $\bm{\mathrm{S}}$ are adequately described in the SERF regime by a polarization vector model where the $x$-axis component of $\bm{\mathrm{S}}$ under steady-state pump-probe and a transverse rf field of angular frequency $\omega$ is given by\cite{BudkerOpticalMagnetometry}
\begin{equation}
S_x = \frac{g_S R_{\mathrm{op}}T_{2,S}^2}{2q^2}\Bigl[\frac{\cos \omega t + (\omega-\omega_S)T_{2,S}\sin\omega t}{1+(\omega-\omega_S)^2T_{2,S}^2}\Bigr]B_1. \label{eq:Sx}
\end{equation}
Here, $\omega_S = g_SB_{z,S}/q$ is the Larmor frequency, $g_S$ is the gyromagnetic ratio, $q$ is the nuclear slowing down factor, $R_{\mathrm{op}}$ is the optical pumping rate and $T_{2,S}^{-1}$ is the transverse relaxation rate of the alkali atom ensemble.  According to the above \autoref{eq:Sx}, the atomic response to $B_1$ is strongest for matched precession frequencies of the spin species: $\omega_I = \omega_S$.  Thus the OPM is tunable to a given NMR frequency $\omega_I$ by setting the magnetic field at the atoms to $B_{z,S}=\pm(q/g_s)\omega_I=\pm\gamma_I(q/g_s)B_{z,I}$.  This adjustment is permitted since $B_{z,I}$ is the superposition of the fields in the interior of coils S2\,+\,S3\,+\,S4, while $B_{z,S}$ is the superposition of fields from S4 and the much weaker exterior field of S2\,+\,S3.  

For Larmor frequencies $\omega_I/(2\pi)$ between 10 and 200 Hz the magnetometer noise is below $10\,\mathrm{fT}/\sqrt{\mathrm{Hz}}$ (\autoref{fig:OPMresponse}b), limited by noise in the lasers and to a lesser extent the Johnson noise of the coils S1\,+\,S2\,+\,S3.  The spin projection noise estimated from the atom density $n_S \approx 10^{20}$ m$^{-3}$, temperature \SI{150}{\celsius} and coherence time $T_{2,S} \approx 3$ ms is $\sqrt{n_S g_S^2 T_{2,S}/q} \sim$ \SI{1.1}{\femto\tesla\per\sqrt\hertz}.  Above fields $B_{z,S} \approx 100$ \si{\nano\tesla}, $\omega_S$ starts to become comparable to $1/T_{2,S}$, marking the limit of the SERF regime, and the magnetometer noise rises above \SI{20}{\femto\tesla\per\sqrt\hertz}.  Overall, as \autoref{fig:OPMresponse}c illustrates, NMR signals are obtainable at fields where Larmor frequencies are around 100 times higher than the atomic bandwidth.  In contrast, without tuning, the combined atomic and nuclear spin system yields a relatively narrow operating range for NMR, quantified by the half-width at half-height of \autoref{eq:Sx}: $\Delta \omega_{I}/(2\pi) \approx g_ST_{2,S}/(2\pi q) \approx$ \SI{80}{\hertz}.  

\subsection*{Dissolved paramagnetic species in liquids}
 
\begin{figure}[ht]
	\includegraphics[width=0.90\columnwidth]{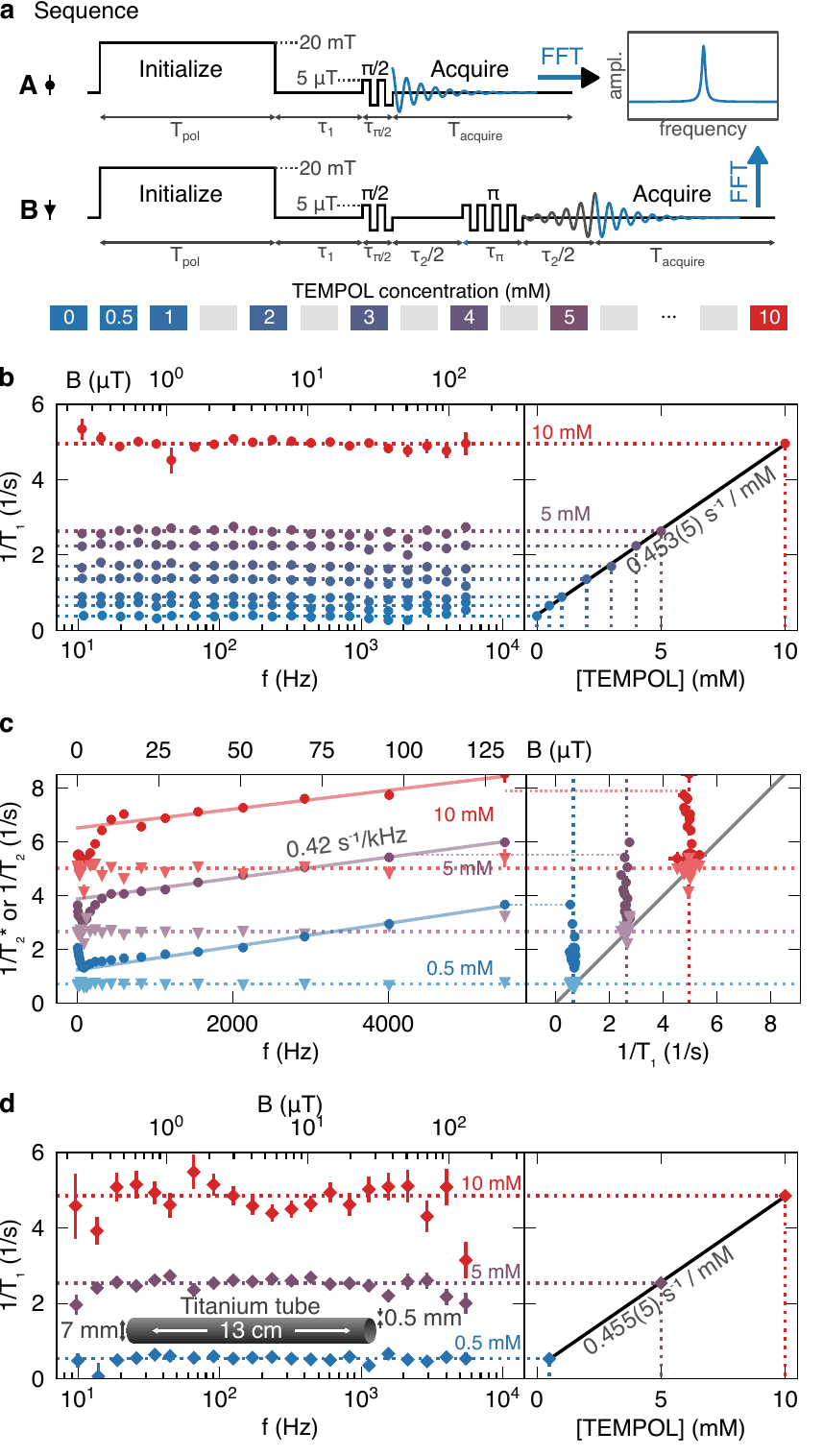}
	\caption{Ultralow-field NMR relaxation of $I$ = \textsuperscript{1}H in aqueous TEMPOL solutions at $30$ $^\circ$C: (a) Sequences A and B measure longitudinal ($T_{1,I}^{-1}$; diamond plot markers) and transverse ($T_{2,I}^{-1}$ and $(T_{2,I}^\ast)^{-1}$; triangle and circle plot markers) relaxation rates.  Vertical axis (not to scale) shows field strength with the polarization field along z and the $\pi/2$ and $\pi$ pulses along y; (b Field dependence of rates $T_{1,I}^{-1}$ across \SI{25}{\nano\tesla} $< B_{z,I} <$ \SI{130}{\micro\tesla}, 1.8 mL sample volume. Rates are linearly proportional to TEMPOL concentration; (c) Field dependence of $(T_{2,I}^\ast)^{-1}$ and $(T_{2,I})^{-1}$.  $(T_{2,I}^\ast)^{-1}$ depends weakly on field due to instrument-specific gradients in $B_{z,I}$ and $B_{z,S}$, while $T_{2,I}^{-1}\approx T_{1,I}^{-1}$; (d) Relaxation rates for TEMPOL solutions encased in 0.5 mm thick titanium tube, 0.1 mL sample volume.  All vertical error bars represent estimated standard deviation errors obtained from least-squares regression analysis.} \label{fig:TEMPOL}
\end{figure}

Many single-component liquids and simple solutions are characterized by an exponential correlation function for molecular tumbling, with a time constant $\tau_{\rm c}$ in the low \si{\pico\second} range.  Unless much slower additional motion processes exist the NMR relaxation times $T_{1,I}$ and $T_{2,I}$ are independent of magnetic field below $B_{z,I} \ll (\gamma_I\tau_{\rm c})^{-1}$) $\sim$~\SI{0.1}{\tesla}, all the way down to ultralow field.

Here we observe the dependence of relaxation in aqueous solutions of the paramagnetic compound 4-hydroxy-2,2,6,6-tetramethylpiperidin-1-oxyl (TEMPOL).  TEMPOL is a chemical oxidant under study elsewhere for potential therapeutic properties\cite{Lewandowski2017IntJMolSci18}, as well as a source of nuclear spin hyperpolarization\cite{Prandolini2009JACS131,Neugebauer2013PCCP15} that can achieve enhanced sensitivity in NMR.   Sequences A and B are used, respectively, to measure \textsuperscript{1}H $T_{1,I}$ and $T_{2,I}$ (\autoref{fig:TEMPOL}a).  

In sequence A, nuclear spin prepolarization at \SI{20}{\milli\tesla} is followed by switching to a lower magnetic field for a time $\tau_1$, before a dc $\pi/2$ pulse induces NMR free nuclear precession about the $z$ axis.  The amplitudes of the NMR signal, $s_A$, are fit well by the function $s_A \propto \exp(-\tau_1/T_{1,I})$ and the observed relaxation rates scale linearly with concentration of the paramagnetic dopant as $T_{1,I}^{-1} = (T_{1,I}^{(0)})^{-1} + k_1[\mathrm{TEMPOL}]$, where $T_{1,I}^{(0)}$ is the relaxation time at zero solute (\autoref{fig:TEMPOL}b).  The relaxivity parameter $k_1 =$ \SI{0.453(5)}{\per\second\per\milli\mole\deci\meter\cubed} is in good agreement with literature values at the high-field end\cite{Prandolini2009JACS131,Neugebauer2013PCCP15}, which gives confidence in the method.  
In sequence B, the initial $\pi/2$ pulse is followed by a Hahn-echo to refocus transverse magnetization after time $\tau_2$.  Signal amplitudes are fit well by the expected function $\exp(-\tau_1/T_{1,I} - \tau_2/T_{2,I})$ and provide a transverse relaxivity parameter $k_2 = 0.455(12)$ \si{\per\second\per\milli\mole\per\deci\meter\cubed} defined by $T_{2,I}^{-1}(B_{z,I} ,[\mathrm{TEMPOL}]) = T_{2,I}^{-1}(B_{z,I},0) + k_2(B_{z,I})[\mathrm{TEMPOL}]$ (\autoref{fig:TEMPOL}c, triangle plot markers).  The result $k_2 = k_1$ holds down to nT fields, which confirms isotropic molecular tumbling in the fast motion limit and absence of slow motional correlations.    

The transverse decay rates are also well approximated by $(T_{2,I}^{\ast})^{-1} = $ FWHM/$(2\pi)$ obtained from line widths in the Fourier-transform NMR spectra of sequence A, due to relatively low inhomogeneity in $B_{z,I}$.  Here,  S2+S3 produce field gradients $\mathrm{d}B_{z,I}/\mathrm{d}z$ and smaller components along $x$ and $y$ due to tilt imperfections in the coil windings, resulting in a linear dependence  $(\mathrm{d}/\mathrm{d}B_{z,I})(T_{2,I}^{\ast})^{-1} =$ \SI{0.01}{\per\second\per\micro\tesla} (or 4 ppk of $B_{z,I}$) observed above 500 Hz \textsuperscript{1}H frequency.  We also note that S4 is centered on the \textsuperscript{87}Rb cell and not on the NMR sample, therefore gradients may cancel out at some Larmor frequencies; this effect is attributed to the line narrowing at around 200 Hz.  Overall the results show that the NMR line width stays below 1 Hz even up to geomagnetic fields, and that TEMPOL causes no further detriment to spectroscopic resolution in the zero/ultralow-field range.

To further demonstrate the application potential of the technique we highlight that ultralow-field cycling and NMR detection is compatible with metal sample enclosures.  NMR signals can be detected without amplitude loss up to \SI{}{\kilo\hertz} Larmor frequencies when \SI{0.1}{\milli\liter} aliquots of the TEMPOL solutions are contained inside a titanium alloy tube (outer diameter \SI{8}{\milli\meter}, inner diameter \SI{7}{\milli\meter}, pressure rating \SI{13}{\mega\pascal}).  Relaxation rates $1/T_{1,I}$ for samples with and without the metal tube, shown in \autoref{fig:TEMPOL}d, are identical within measurement error to those of \autoref{fig:TEMPOL}b.  Larger error bars are due to the smaller sample volume giving lower snr.  This measurement is impossible via conventional fast-field cycling NMR techniques, where eddy currents in metal strongly attenuate the amplitude of high-frequency NMR signal and also limit the rate of field switching.  Transverse relaxation rates are also unaffected by the presence of the metal tube, from which we may conclude that eddy currents are negligible over the relatively small (mT) range of field switching.  The approach may therefore open the way to study relaxation in new contexts, for instance high-pressure fluids (e.g.\ supercritical fluids), flow in pipes, foil-sealed products (e.g.\ foods, pharmaceuticals) and (e.g.\ lead-, tungsten-) sealed radioactive samples.

\subsection*{Liquids confined in porous materials}
To demonstrate new insight into molecular motion near pore surfaces we study the $^1$H spin relaxation of n-alkane hydrocarbons confined within matrices of alumina ($\gamma$ polymorph, \SI{9}{\nano\meter} mean pore diameter) and titania (anatase polymorph, 7--10\,\si{\nano\meter} mean pore diameter).  These simple inorganic oxides in their mesoporous form possess catalytic features due to their high specific surface area, Lewis acidic sites and option of chemical treatments including metalization to activate the pore surface.  Yet, owing to the frequency range of conventional NMRD techniques, there is limited understanding of how molecular dynamics and surface site properties relate to long-$\tau_{\rm c}$ relaxation processes, even without surface functionalization\cite{WardWilliams2018JPCC122}.

\begin{figure}[t]
	\includegraphics[width=\columnwidth]{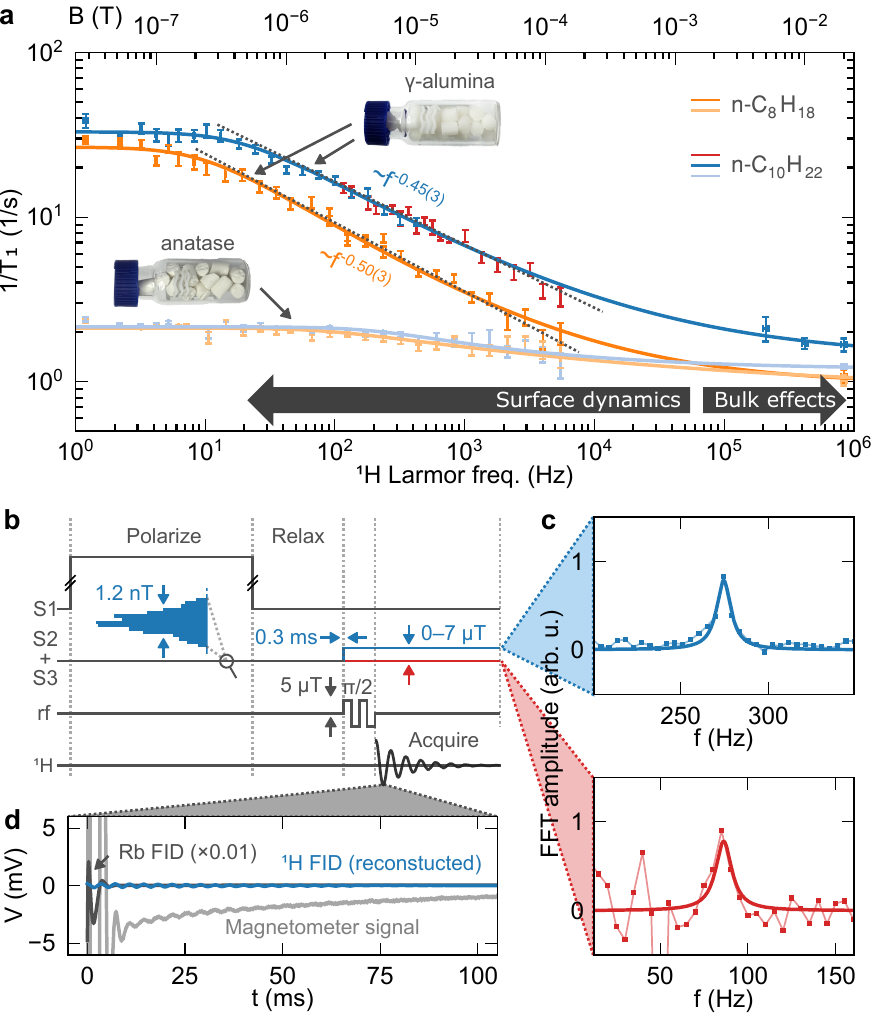}
	\caption{Ultralow-field FFC NMR of pore-confined fluids: (a) Magnetization decay rates for \textit{n}-octane and \textit{n}-decane in porous $\gamma$-alumina and anatase titania.  For \textit{n}-decane in $\gamma$-alumina, colors indicate the NMR detection field, as discussed in the main text and represented in the rest of the figure.  Data above 100 kHz correspond to magnetization buildup rates in the S1 field; vertical error bars represent estimated standard deviation errors obtained from least-squares regression analysis; (b) FFC sequence used to measure $T_{1,I}$ decay at ultralow field; (c) frequency- and (d) time-domain NMR signal of \textit{n}-decane in $\gamma$-alumina after relaxation at $\omega_I/(2\pi)=85$ \si{\hertz}.  In (c), the NMR signal-to-noise ratios (snr) illustrate the FFC requirement below 100 Hz: poor snr at $\omega_I/(2\pi)=85$ \si{\hertz} is due to $1/f$ and 50 Hz noise of the OPM; FFC switching to $\omega_I/(2\pi) =$ \SI{275}{\hertz} between relaxation and detection events is optimal for high snr and short dead time.  The blue and red color coding also serves to indicate the method used to measure $T_{1,I}$ for \textit{n}-decane in $\gamma$-alumina, in part (a) of the Figure.  The reconstructed time-domain signal in (d, blue curve) equates to the Lorentzian line shape fitted in (c).
	}
	\label{fig:pores}
\end{figure}

\autoref{fig:pores}a shows \textsuperscript{1}H relaxation rates at 30\,$^\circ$C for imbibed \textit{n}-alkanes, measured between 1\,Hz and 5.5\,kHz Larmor frequency using the sequence shown in \autoref{fig:pores}b.  Due to excess noise in the magnetometer below 100 Hz (including mains electricity noise and $1/f$ noise, see \autoref{fig:pores}c), fast field switching between relaxation and detection events is the preferred measurement option to probe the lowest fields, where the NMR signal is always detected at frequency above 100 Hz.  Above 100 Hz Larmor frequency, the noise floor is low enough to detect NMR signals at the relaxation field, without switching.  The measurable NMR relaxation is limited in principle to rates $T_{1,I}^{-1} < R_{\mathrm{op}}^{-1}$, where the latter is of order \SI{300}{\per\second}.  However, in practice, the limit is $T_{1,I}^{-1} < T_{2,S}^{-1}$ or around \SI{100}{\per\second} since the atomic precession signal causes a \SI{10}{\milli\second} dead time caused following the $\pi/2$ pulse (see \autoref{fig:pores}d).  

The main feature of \autoref{fig:pores}a is the weak dispersion in $T_{1,I}^{-1}$ for each alkane, and moreover between the two porous materials, across the conventional FFC-NMR frequency range 10 kHz to 1 MHz\cite{WardWilliams2018JPCC122}.  The relaxation rate for each alkane depends only slightly on the porous material, therefore bulk effects dominate the relaxation process in this range.  In contrast, relaxation rates below 10 kHz depend strongly on the material and mechanisms related to the surface are prominent, with the higher values being observed towards zero field.  The $T_{1,I}$ dispersion in titania is much weaker than in $\gamma$-alumina; $T_{1,I}^{-1}$ reaches only around \SI{2}{\per\second} below 200 Hz, compared to \SI{30}{\per\second} for alumina.  Although the two materials have similar mean pore diameter and surface area/volume ratio, surface-induced relaxation is not so active in the first material.  It is known from electron spin resonance spectroscopy\cite{JWW2021JOCC125} that the alumina contains a higher concentration of paramagnetic impurity -- [Fe\textsuperscript{3+}] $\approx$ \SI{2e16}{\per\gram} (i.e., ions per unit mass of the dry porous material) in alumina vs.\ \SI{2e15}{\per\gram} in titania --  suggesting that the lower-frequency relaxation mechanism involves dipole-dipole coupling between \textsuperscript{1}H and the surface spins, rather than surface-induced modulation of intra-molecular \textsuperscript{1}H-\textsuperscript{1}H spin couplings.

Between $\omega_I/(2\pi)=50$ Hz and 5000 Hz, the longitudinal relaxation in $\gamma$-alumina obeys a power-law frequency dependence: $T_{1,I} \propto \omega_I^\xi$.  Fitted slopes $-\mathrm{d}(\mathrm{log}\,T_{1,I}^{-1})/\mathrm{d}(\mathrm{log}\,\omega_I)$ give exponents $\xi = 0.50\pm0.03$ for octane and $\xi = 0.45\pm0.03$ for decane.  Such values are consistent with simple numerical simulations in which imbibed molecules randomly walk within a dilute matrix of non-mobile spins -- such as surface paramagnets -- where the strength of dipole-dipole interactions between the two spin species scales with the inverse cube of their instantaneous separation\cite{Mcdonald2018PRE98}.  This nonlinear dependence results in an example of L\'{e}vy walk statistics.  A detailed characterization of these effects in the alumina system is ongoing work.  

Although developing an analytical model for the surface dynamics is outside the scope of this paper, for analysis of the correlation time it suffices to fit the measured relaxation rates by a stretched Lorentzian function $[T_{1,I,\mathrm{fit}}(\omega_I)]^{-1} = [T_{1,I}(0)]^{-1}/(1+\tau_\mathrm{c}^2\omega_I^2)^{\beta} + [T_{1,I}(\infty)]^{-1}$ with four independent fit parameters: $T_{1,I}(0)$, $T_{1,I}(\infty)$, $\tau_\mathrm{c}$ and $\beta$.  For $\omega_I\tau_\mathrm{c} \gg 1$ and $T_{1,I,\mathrm{fit}}(\omega) \ll T_{1,I}(\infty)$, the function is approximated by a power law with $\xi = 2\beta$.  The fitted curves are plotted as solid lines in \autoref{fig:pores}a.  The parameter $\tau_c$ for alkanes in alumina is determined from the relaxation behavior below 50 Hz, where $T_{1,I}(\omega_I)$ changes from a power law frequency dependence to a constant, i.e., towards a plateau at $T_{1,I}(0)$.  Using the analysis presented earlier, this indicates a maximum correlation time ($\tau_c=\tau_{\rm c,max}$) of around 20-30 ms, which is at least two orders of magnitude longer than the maximum correlation time of more polar molecules in porous confinement, such as water.  Relative to octane, the plateau for decane extends to a higher Larmor frequency, indicating a shorter $\tau_{\rm c,max}$,  despite octane having a higher self-diffusion coefficient as a bulk liquid.  
However, at this point $\tau_{\rm c}$ is also of similar magnitude to the longitudinal relaxation time.  Under such conditions the assumptions of standard NMR relaxation theories -- such as the Wangsness-Bloch-Redfield theory -- are not strictly justified as valid, in particular the coarse graining of time\cite{Redfieldtheory}, where spin diffusion may be a part of the relaxation mechanism, or set an upper limit for the relaxation rate in the plateau.  Whether this is true requires more information on the physical process responsible for spin relaxation.  

\subsection*{High-resolution relaxometry}

\begin{figure*}[t]
	\includegraphics[width=\textwidth]{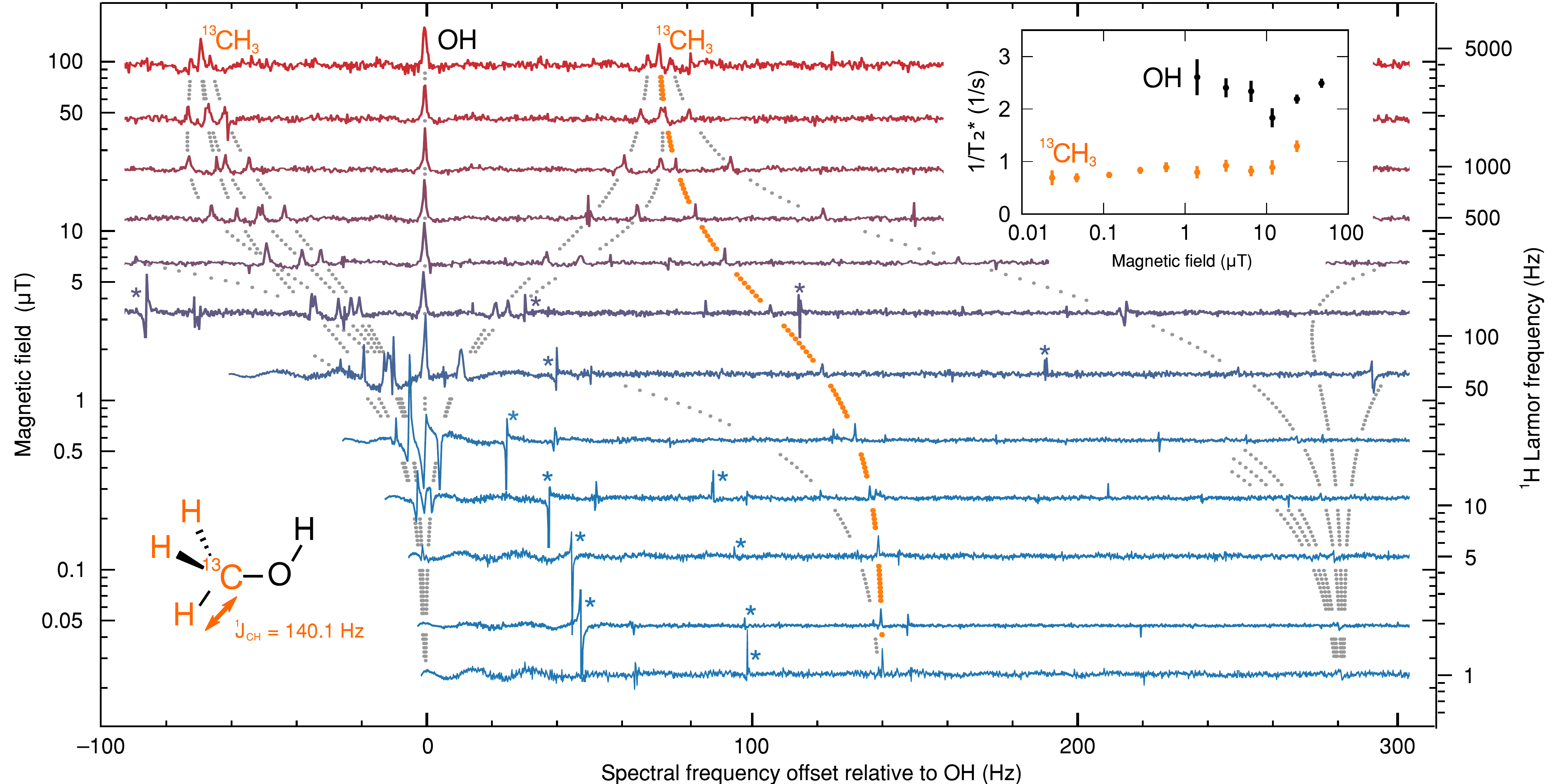}
	\caption{
	NMR spectra of \textsuperscript{13}CH\textsubscript{3}OH versus field strength in the range \SI{25}{\nano\tesla} to \SI{100}{\micro\tesla}.  Horizontal lines show NMR spectra acquired as in \autoref{fig:OPMresponse}, offset vertically by the field at which they were acquired (left scale), and horizontally so as to place the uncoupled OH resonance at zero.  Individual resonances for the chemically distinct CH\textsubscript{3} and OH spin groups are clearly visible and demonstrate relative frequency shifts due to the transition between strong and weak heteronuclear coupling regimes corresponding to low and high field, respectively.  The different peaks correspond to different spin combinations; in high field the peaks correspond to single-spin transitions of \textsuperscript{1}H, while at zero field peaks at $^1 J_{\rm CH} \,=$ \SI{140.1}{\hertz} and $2\times ^1 J_{\rm CH} \,=$ \SI{280.2}{\hertz} correspond respectively to singlet-to-triplet and triplet-to-quintet transitions of the \textsuperscript{13}CH\textsubscript{3}.  Features marked with an asterisk ($\ast$), e.g.\ \SI{50}{\nano\tesla} and $\approx \SI{48}{\hertz}$ plus $\approx \SI{98}{\hertz}$, are artefacts due to \SI{50}{\hertz} line noise and harmonics. Dotted curves show the predicted resonance frequencies as a result of the heteronuclear coupling. Inset shows transverse decay rate $1/T_{2,I}^\ast$ for the CH\textsubscript{3} signal peak (dotted orange curve) and OH signal peak (at \SI{0}{\hertz}) versus field strength. Error bars represent estimated standard deviation errors obtained from least-squares regression analysis.} \label{fig:MeOH}
\end{figure*}

Field instability in traditional NMRD electromagnets is large compared to spectroscopic dispersion from NMR chemical shifts or inter-spin couplings, resulting in severe or even complete overlap of the signals from nuclei in different chemical groups or different compounds in a mixture.  Although additional strategies may prove helpful to assign relaxation rates to distinct chemical groups, (e.g.\ selective deuteration or other isotopic substitution, inverse Laplace transforms), higher-resolution signal detection would be a more general and direct solution.

Here we illustrate simultaneous measurement and independent fitting of relaxation rates for two chemically distinct \textsuperscript{1}H environments in methanol (CH\textsubscript{3}OH).  A scalar coupling ($^1 J_{\rm CH} \,=$ \SI{140.1}{\hertz}) between \textsuperscript{13}C and \textsuperscript{1}H nuclei in the \textsuperscript{13}CH\textsubscript{3} group shifts the corresponding NMR signal by around $\pm 0.5 J_{\rm CH}$ relative to that of the non-coupled OH, when measured at fields $|B_{z,\rm I}| \gg |2\pi ^1J_{\rm CH}/(\gamma_{\rm H}-\gamma_{\rm C})|$.  The latter criterion defines the well-known weak heteronuclear coupling regime.  Shifts by other multiples of $^1J_{\rm CH}$ between 1 and 2 occur at lower fields.  Experimental spectra and simulated positions of the NMR peaks are shown in \autoref{fig:MeOH}.  Line widths are on the order of 1 Hz, which should also allow spectral resolution of the CH\textsubscript{3} groups in methanol, acetone (CH\textsubscript{3}COCH\textsubscript{3}, $^1 J_{\rm CH} \,=$ \SI{127}{\hertz}), acetic acid (CH\textsubscript{3}COOH, $^1 J_{\rm CH} \,=$ \SI{130}{\hertz})  dimethylsulfoxide (CH\textsubscript{3}SOCH\textsubscript{3}, $^1 J_{\rm CH} \,=$ \SI{137}{\hertz}) and other solvents. Isotopomers splittings that arise for couplings over more than one chemical bond, e.g.\ --\textsuperscript{13}C\textsuperscript{12}CH\textsubscript{3}, for which  $^2 J_{\rm CH} \,=$ 5--30 \si{\hertz}, would also be resolvable.

By using the sequence shown in \autoref{fig:TEMPOL}a to provide a series of $T_{1,I}$-weighted spectra, a fitted relaxation rate $1/T_{1,{1\rm H}}$ = \SI{0.44(5)}{\per\second} (not plotted) is obtained for the OH subsystem.  Within error, the value does not depend on field.  Relaxation rates for the CH\textsubscript{3} subsystem are also field independent in the weak-coupling regime above $B_{z,\rm I} =\SI{10}{\micro\tesla}$, and are very close to those of OH: $1/T_{1,{1\rm H}}$ =  \SI{0.45(14)}{\per\second} (not plotted).  Both sets of rates refer to relaxation of the \textsuperscript{1}H spin species and are thus comparable with results obtained via conventional field cycling NMRD.

At lower fields, however, L-S type effects of the scalar coupling between \textsuperscript{13}C and \textsuperscript{1}H\textsubscript{3} lead to a significant contrast in transverse relaxation rates.  Of most interest is the peak tending to frequency $^1J_{\rm CH}$ at zero field (orange curve in \autoref{fig:MeOH}), which corresponds to singlet-to-triplet coherence in the isolated manifold formed between \textsuperscript{13}C and the \textsuperscript{1}H\textsubscript{3} state of total spin quantum number 1/2.  Here the transverse relaxation rate is around 2.5 to 3 times slower than for the non-coupled OH (\autoref{fig:MeOH}, inset).  The \textsuperscript{13}CH\textsubscript{3} system therefore exhibits a type of long-lived spin order in the NMR ensemble\cite{LLSObook}.  Despite the presence of \textsuperscript{1}H--\textsuperscript{1}H and \textsuperscript{13}C--\textsuperscript{1}H dipole-dipole couplings, this singlet-to-triplet coherence is long lived because it is less sensitive to relaxation by fields that are correlated across the \textsuperscript{13}C and  \textsuperscript{1}H spin groups. This includes much of the intra-\textsuperscript{13}CH\textsubscript{3} dipole coupling as well as longer-range couplings.  The result has potential importance in applications to probe dipole-dipole interactions at short distances away from the CH\textsubscript{3}, including intermolecular interactions.

\section*{Discussion and outlook}
The study demonstrates that unique, important information about nuclear spin relaxation in liquids can be obtained by fast field switching and tunable NMR detection at ultralow magnetic fields.

A basic advantage of the low-field NMR detection is that it eliminates many concerns about magnetic field homogeneity and stability, since magnetic fields are accurately and precisely controlled.  This is shown by the result $T_2 \sim T_1$ for the series of TEMPOL solutions.  Additionally, as shown for methanol, the Fourier-transform NMR spectrum line width is adequate to resolve spin-spin couplings (even in the alumina system where line widths exceed, \SI{10}{\hertz}) and therefore components of liquid mixtures. Ultralow-field NMRD may therefore be able to probe other interplay, such as competitive adsorption between molecules and pore. 
Besides nonuniformity of applied magnetic fields, conventional NMR is also confounded by sample heterogeneity, especially in multi-phase samples with internal magnetic susceptibility variation, including porous materials, or metal regions.  The $\tau_{\rm c}$ values obtained here for alkanes in porous alumina are extremely long by FFC-NMR standards - comparable $\tau_{\rm c}$s are typically probed in high field by pulsed-field gradient (PFG) diffusometry and rotating-frame ($T_{1\rho,I}$) relaxometry techniques\cite{Pricebook}.  In most if not all applications, both of the latter are highly susceptible to contamination by poor field homogeneity and radiofrequency offset errors.

Compared to high-field inductive-detected NMRD, ultralow-field OPM-detected NMRD currently has some limitations.  A main limitation, resulting from the Hartmann-Hahn matching condition, is that the OPM Faraday rotation signal contains free-precession responses of the sensor atom and NMR sample spins at the same frequency.  The atomic response is at least two orders of magnitude stronger than the NMR signal and can easily saturate the digitizer, which leads to a ``dead time'' on the order of the optical pumping time (10 ms, see \autoref{fig:pores}d).  This currently hinders applications in chemical systems where molecules interact more strongly, namely liquids in nanopores (e.g.\ zeolites, shale) and interfaces with hydrogen bonding, where relaxation rates are higher.  In principle, Q-switching of the optical pumping beam\cite{Limes2020PRApplied14} is a method to accelerate magnetometer recovery after the magnetic field pulses and reduce the dead time down to the field switching time, well below 1 ms, without compromising sensitivity. 

Ultralow-field FFC NMRD may also in the future offer new study paths when enriched by nuclear spin hyperpolarization.  As shown in \autoref{fig:OPMresponse}c, a few tens of scans result in snr $>$20 dB, even though the spins are only prepolarized to around 1 part in $10^8$ at the \SI{20}{\milli\tesla} starting field.  Nitroxide radical compounds such as TEMPOL are a source of higher electron spin polarization, around 1 part in $10^5$ at \SI{20}{\milli\tesla}, that can be efficiently transferred to nuclei via the Overhauser effect at both high\cite{Prandolini2009JACS131} and ultralow\cite{Hilschenz2019JMR138} magnetic fields.  Hyperpolarization via surface-supported paramagnetic species and other spin transfer catalysts may also be an option to study nuclear polarization buildup near pore surfaces, providing information that may differ from relaxation decay.  TEMPOL and other persistent radicals are used to prepare hyperpolarized biochemical probes for clinically relevant in-vivo observations of disease via MRI\cite{Sriram2014emagres3}.  These systems could profit from a knowledge of signal decay mechanisms at ultralow magnetic fields, for new sources of image contrast or to minimize polarization losses before imaging/detection.

\section*{Methods}
\subsection{Sample preparation}

All samples studied in this work were contained in disposable glass vials (12\,mm o.d., 20\,mm length, 1.8\,mL internal volume, 8--425\,thread) sealed with a silicone septum and finger-tight polypropylene screw cap.  

\textit{Preparation of TEMPOL samples:} a 10 mM stock solution of the radical 4-hydroxy-2,2,6,6-tetramethylpiperidin-1-oxyl (Sigma Aldrich, CAS: 2226-96-2) was prepared in 5.0 mL deoxygenated milli-Q water and diluted to concentrations of 0.5, 1, 2, 3, 4, 5 and 10 mM with deoxygenated milli-Q water.  The diluted solutions were not further de-gassed. 

\textit{Preparation of porous materials samples:} cylindrical extrudate pellets of meso-porous $\gamma$-alumina (Alfa Aesar product 43855, lot Y04D039: 3\,mm diameter, 3\,mm length, 9 nm BJH mean pore diameter, Langmuir surface area \SI{250}{\meter\squared\per\gram}) and anatase titania (Alfa Aesar product 44429, lot Z05D026: 3\,mm diameter, 4\,mm length, 7--10\,nm mean pore size, Langmuir surface area \SI{150}{\meter\squared\per\gram}) were obtained commercially.  Pellets were oven dried at 120\,$^\circ$C for 12h to remove physisorbed H$_2$O and then imbibed in neat n-alkane for at least 12 h after recording the dry mass.  Excess liquid on the pellet outer surface was gently removed using tissue paper.  The pellets were then placed in a vial (see \autoref{fig:pores}a), sealed with the cap and the combined mass of pellet and imbibed hydrocarbon was recorded. 

\textit{Preparation of methanol sample:} 0.9 g of \textsuperscript{13}C-methanol (\textsuperscript{13}CH\textsubscript{3}OH 99\%, Sigma Aldrich product 277177) was added to the sample vial without dilution, then followed by N\textsubscript{2} bubbling (2-3 min) to displace dissolved paramagnetic O\textsubscript{2}.  

\subsection{Optical magnetometer}
The magnetometer used to detect \textsuperscript{1}H precession signals in the NMR samples was operated as follows.  A cuboid borosilicate glass cell of inner dimensions $5 \times 5 \times 8\,\mathrm{mm}^3$ contained a droplet of rubidium-87 metal and \SI{90}{\kilo\pascal} N\textsubscript{2} buffer gas (Twinleaf LLC).  The cell was electrically heated to 150 $^\circ$C to vaporize the alkali metal.  A circularly polarized light beam along $z$ axis (3\,mW, tuned to the center of the collision-shifted D\textsubscript{1} wavelength) optically pumped the atomic spin polarization to $S_z \approx 0.5$.  Faraday rotation in a second, linearly polarized light beam (10\,mW, \SI{65}{\giga\hertz} red-shifted from the pump, along $x$ axis) was used to non-resonantly probe the $S_x$ component of atomic polarization.  The probe beam was linearly polarized and slightly detuned from the \textsuperscript{87}Rb D$_1$ transition, such that on passing through the cell along the $x$ axis its axis of polarization was optically rotated by an angle proportional to $S_x$.  The Faraday rotation was detected by polarimetry using a differential photodetector (Thorlabs PDB210A), which produced an analog voltage signal that was conditioned (amplified, filtered to eliminate high-frequency noise and dc offset) and digitized (60 ksps, 16-bit $\pm 5\,\mathrm{V}$ ADC range) before storage and further processing on a computer.    

\subsection{Magnetic coils and shielding}
The vapor cell and heating assembly was placed as close as possible to the NMR sample at a standoff distance $d_1 = 3.5\,\mathrm{mm}$ between outer walls of the vial and atomic vapor cell (see \autoref{fig:OPMresponse}a).  In order of increasing distance away from the NMR sample, $d_1$ accounts for (i) S2\,+\,S3 coil windings (34 AWG enameled copper wire, solenoid length 13\,cm, diameter 14\,mm), (ii) a carbon-fiber support structure, (iii) S1 coil windings (36 AWG enameled copper wire, solenoid length 2.5\,cm, single layer), (iv) a water-cooling jacket (de-ionized water, flow rate \SI{1}{\milli\liter\per\second}) to remove heat deposited when the polarizing coil is energized and to maintain a stable sample temperature, (v) a PEEK support structure and (vi) an air gap for further thermal insulation.  The entire structure was operated within a cylindrical magnetic shield (Twinleaf LLC, model MS-1F) of 20 cm outer diameter and length 30 cm.  The main axis of the cylinder was co-axial with the pump beam axis.

Field-to-current ratios inside each coil were calibrated using the frequency of \textsuperscript{1}H precession frequency in de-ionized water.  These were S2: \SI{7.59(3)}{\micro\tesla\per\milli\ampere}; S3: \SI{7.50(3)}{\micro\tesla\per\milli\ampere}; S4: \SI{150.1(5)}{\nano\tesla\per\milli\ampere}.  The atomic spin precession frequency at the atomic vapor cell was used to calibrate the external field of coils S2 \SI{-11.1(3)}{\nano\tesla\per\milli\ampere} and S3: \SI{-4.4(2)}{\nano\tesla\per\milli\ampere}, which equate to \SI{1460}{ppm} and \SI{580}{ppm} of the field at the vial, respectively. 

\subsection{Magnetic noise spectra}
An ac test signal of $\pm8$\,\si{\pico\tesla} along the \textit{y} axis was applied to calibrate the Faraday response as a function of frequency and \textit{z} bias field.  The calibration vs.\ frequency was used to scale the spectral response of the balanced photodetector from units of \si{\volt\per\sqrt\hertz} into \si{\tesla\per\sqrt\hertz}.  The maximum magnetic response at a given bias field was confirmed to equal the tuning condition where the atomic Larmor frequency matched the frequency of the ac signal, given the prior calibration of the magnetic field at the sample vial and magnetometer cell.  

\subsection{Field switching}
Timing of the NMR pulse sequences and data acquisition was controlled by a microcontroller (Kinetis K20 series: time base \SI{2}{\micro\second}, precision \SI{17}{\nano\second}, CPU speed 120 MHz).  Current to the polarizing coil was switched via a dual H-bridge circuit with parallel flyback diodes and the switching time was $<1.0$\,\si{\milli\second}.  Coils S2 and S4 were connected to a low-noise precision current source (Twinleaf model CSB-10, 20-bit resolution over $\pm10$\,\si{\milli\ampere}) with a low-pass LC filter in series, resulting in a combined switching and settling time of order 100 ms.  The FFC solenoid coil S3 operated at a current $<1\,\mathrm{mA}$ direct from the microcontroller digital-to-analog converter (12-bit resolution, 0-1\,\si{\milli\ampere}) for rapid and precise field switching without feedback controls.  Typical S3 switching times were 0.25\,\si{\milli\second} and the accuracy (determined from standard error in the mean NMR center frequency over repeated scans at $\omega_I/(2\pi)=550$\,\si{\hertz}, see \autoref{fig:pores}b inset) was around 1\,\si{\nano\tesla}.  The residual interior field of the MuMetal shield along the $x$, $y$, and $z$ axes of $\sim$~\SI{10}{\nano\tesla} was also compensated for.  

Under steady-state conditions with the pre-polarizing coil turned off, the cooling system maintained a temperature of 27--28\,$^\circ$C at a thermocouple attached to the outside wall of the sample vial.  The steady-state temperature rose to 30--31\,$^\circ$C when the polarizing coil was energized at 20\,mT (2.2\,A).

\section*{Acknowledgments}
The work described is funded by: 
EU H2020 Marie Sk{\l}odowska-Curie Actions project ITN ZULF-NMR  (Grant Agreement No. 766402); 
Spanish MINECO projects OCARINA (Grant No. PGC2018-097056-B-I00), the Severo Ochoa program (Grant No. SEV-2015-0522); 
Generalitat de Catalunya through the CERCA program; 
Ag\`{e}ncia de Gesti\'{o} d'Ajuts Universitaris i de Recerca Grant No. 2017-SGR-1354;  Secretaria d'Universitats i Recerca del Departament d'Empresa i Coneixement de la Generalitat de Catalunya, co-funded by the European Union Regional Development Fund within the ERDF Operational Program of Catalunya (project QuantumCat, ref. 001-P-001644); Fundaci\'{o} Privada Cellex; Fundaci\'{o} Mir-Puig; MCD Tayler acknowledges financial support through the Junior Leader Postdoctoral Fellowship Programme from ``La Caixa'' Banking Foundation (project LCF/BQ/PI19/11690021).  The authors also thank Jordan Ward-Williams and Lynn Gladden (University of Cambridge) for providing samples of porous alumina and titania, and for discussions. 

\section*{Author contributions}
MCD Tayler proposed the study.  S Bodenstedt prepared the samples, measured and analyzed the experimental data and together with MCD Tayler built the experimental apparatus and made the theoretical interpretation.  MCD Tayler wrote the manuscript with input from all authors.  All authors reviewed the manuscript and suggested improvements. MCD Tayler and MW Mitchell supervised the overall research effort. 

\section*{Data availability}
The raw data generated in this study have been deposited in the OpenAIRE database under accession code \url{https://doi.org/10.5281/zenodo.4840653}

\section*{Competing interests}
The authors declare no competing interests.

\section*{References}
\normalem

\end{document}